%Paper: nucl-th/9404021
%From: "ekpvs2::gruen"@evax.tuwien.ac.at
%Date: Tue, 19 Apr 1994 12:46:07 +0100

\magnification=\magstep1
\overfullrule=0pt

\def\2spaltig#1#2{\par\noindent
\vbox to 1.5 truecm
{\hsize=7.5 truecm
{\par\noindent#1\vfill}}
\hskip 1 truecm
\vbox to 1.5 true cm
{\hsize=7.5 truecm
{\par\noindent#2\vfill}}}

\centerline{\bf Systematics of $\alpha$-nucleus optical potentials}
\medskip
\centerline{P.~Mohr, H.~Abele, U.~Atzrott, G.~Staudt (Physikalisches Institut,
Univ.~T\"ubingen),}
\centerline{R.~Bieber, K.~Gr\"un, H.~Oberhummer (Institut f\"ur
Kernphysik, TU Wien),}
\centerline{T.~Rauscher (Institut f\"ur Kernchemie, Univ.~Mainz)}
\centerline{E.~Somorjai (ATOMKI, Debrecen)}
\bigskip
\noindent {\bf 1. Introduction}
\medskip
\noindent For the description of nuclear processes in many
astrophysical scenarios the knowledge of $\alpha$-nucleus potentials
is necessary. Such processes are radiative capture, transfer
reactions and alpha-decay occuring in primordial nucleosynthesis,
stellar hydrostatic and explosive burning modes.
\medskip
\noindent Until now such nuclear processes have often been described using
mainly phenomenological and energy-independent potentials (e.g.~square
well, Woods-Saxon potentials etc.). In this work we develop
$\alpha$-nucleus potentials using the folding procedure [1,2]. With
this method the ambiguity of the phenomenological potentials can be
avoided to a great extent. The uniqueness and the energy dependence
of these potentials are an important feature with respect to
astrophysical applications.
\medskip
\noindent Such $\alpha$-nucleus potentials have been used
successfully for the description of scattering processes [3,4,5],
transfer reactions [6,7,8] and radiative capture [9,10] on light
nuclei. In this work we extend our systematic investigation of
$\alpha$-nucleus potentials to intermediate and heavy stable and
unstable nuclei $(A \geq 70)$ which are relevant for the p-process
[11].
\bigskip
\noindent {\bf 2. Folding procedure}
\medskip
\noindent The real part of the optical potential is deduced in the
framework of the double-folding model of Kobos et al.~[1] and is
described by
$$V(r) = \lambda \int d\vec{r}_1 \int d\vec{r}_2 \rho_T(\vec{r}_1)
\rho_\alpha (\vec{r}_2) t (E,\rho_T, \rho_\alpha, \vec{s} = \vec{r} +
\vec{r}_2 - \vec{r}_1)~~, \eqno(1)$$
where $\vec{r}$ is the separation of the centers of mass of the
colliding target nucleus and the $\alpha$ particle,
$\rho_T(\vec{r}_1)$ and $\rho_\alpha(\vec{r}_2)$ are the respective
nucleon densities derived from nuclear charge distributions [12], and
$t(E, \rho_T, \rho_\alpha,s)$ is the density-dependent effective NN
interaction [1]. By means of  the normalization factor $\lambda$ the
depth of the potential can be adjusted to elastic scattering data and
to bound and resonant state energies of nuclear cluster states.
\medskip
\noindent The imaginary part of the potential can be parametrized by
a Woods-Saxon form.
\medskip
\noindent The strength of the potential is measured by its volume
integral per interacting pair of nucleons, e.g.~for the real part
$$J_R(E) = {4\pi \over A_P \cdot A_T} \int^\infty_0
V(r,E)r^2dr~~,\eqno(2)$$
where $A_P$ and $A_T$ denote the projectile and target mass numbers,
respectively.
\topinsert
\vskip 6.2 truecm
\2spaltig{\noindent Fig.~1a: Volume integrals of the real part of the
optical potential for the $\alpha~+~^4$He, $\alpha~+~^{16}$O, and
$\alpha~+~^{40}$Ca system.}
{\noindent Fig.~1b: Volume integrals of the imaginary part of the
optical potential for the $\alpha~+~^4$He, $\alpha~+~^{16}$O, and
$\alpha~+~^{40}$Ca system.}
\endinsert
\noindent In Figs.~1a and 1b the volume integrals for the real part,
$J_R$, and the imaginary part, $J_I$, of the optical $\alpha$-nucleus
potential for the target nuclei $^4$He, $^{16}$O and $^{40}$Ca are
shown. These values have been obtained from the analysis of elastic
scattering data and from calculations of $^8$Be, $^{20}$Ne and
$^{44}$Ti cluster states, respectively [3,4]. A strong energy and
mass dependence for both, the real and imaginary parts, can be
observed. The energy dependence of $J_R$ (Fig.~1a) is due to the
energy dependence of both, the effective NN interaction (Eq.~1) and
the so-called  dynamic polarization potential, which is related  to
that of the imaginary part of the potential by a dispersion relation
[13,3]. The curves shown in Fig.~1a represent the results of
calculations which consider both effects [4]. In these calculations
the energy dependence of $J_I$ was parametrized by [14]
$$\eqalign{J_I &= J_0 {(E - E_0)^2 \over (E-E_0)^2 + \Delta^2} \cr
 J_I &= 0 \cr}
\qquad
\left.\eqalign{E &\geq E_0 \cr
               E &< E_0 \cr}\right\} \eqno(3)$$
with $E_0$ being the threshold energy for inelastic
processes. A linear regression procedure to the data points
results in values for $J_0$ and $\Delta$. The curves calculated with
these parameters are given in Fig.~1b as solid, dashed and dotted
lines.
\pageinsert
\vskip 22 truecm
\noindent Fig.~2: Elastic $\alpha$ scattering on $^{90}$Zr,
$^{208}$Pb, $^{70}$Ge, $^{141}$Pr and $^{144}$Sm: Experimental data
[25-37] and optical model fits calculated by using double-folded
potentials.
\endinsert
\bigskip
\noindent {\bf 3. Alpha-nucleus potentials for intermediate and heavy
nuclei}
\medskip
\noindent Using the same procedure as described in Sec.~2, we
determined optical alpha-nucleus potentials for intermediate and
heavy nuclei. In Fig.~2 calculated elastic scattering cross
sections for some target nuclei are compared with experimental data
at different energies. An excellent agreement between the theoretical
analysis and the experiment is obtained. From this fit to the
experimental data, the strengths of the real part, $\lambda$, as well
as the parameters of the imaginary Woods-Saxon potentials are
obtained for energies above 15 MeV.
\medskip
\noindent  The volume integrals for the real as well as the imaginary
part of these potentials are shown in Figs.~3a and 3b. The following
results can be deduced:
\topinsert
\vskip 6.3 truecm
\2spaltig{Fig.~3a: Volume integrals of the real part of the optical
$\alpha$-nucleus potential for some intermediate and heavy nuclei.}
{Fig.~3b: Volume integrals of the imaginary part of the optical
$\alpha$-nucleus potential for some intermediate and heavy nuclei.}
\endinsert
\smallskip
\item{(i)} The mass dependence of the volume integrals for the real
part of the potentials is very weak for all heavier nuclei.
However, the absolute values are somewhat smaller than those for
lighter nuclei (compare Figs.~1a and 3a).
\smallskip
\item{(ii)} The energy dependence has a similar form as the one
obtained for lighter nuclei. The volume integrals of the real part
have a maximum
about $E_{CM} = 30$ MeV and decrease slightly when going to lower and
higher energies. For astrophysically relevant energies, which are in the
order of 10 MeV, a linear extrapolation to lower energies is
performed (s.~Fig.~3a). We obtain
$$\eqalign{J_R/(A_P \cdot A_T) = (320 &+ 0.67 E_{CM})~\hbox{MeV~fm}^3
\cr
{}~~~~~~~~~~~~&(E_{CM}~\hbox{in~MeV}) \cr}\eqno(4)$$
\item{} The curve shown in Fig.~3a is the result of a calculation
for the $\alpha$-$^{208}$Pb potential which contains the energy
dependence of the folding potential and the dispersive part.
The knowledge of bound and quasibound-state potentials is also
necessary for the calculation of transfer and capture cross sections.
We have calculated these alpha-nucleus potentials for some bound
states. The volume integrals for these potentials are also close to
the extrapolated values (s.~Fig.~3a).
\smallskip
\item{(iii)} The volume integrals for the imaginary part of the
potentials obtained from the fit to the experimental elastic
scattering data for the target nuclei $^{70}$Ge, $^{90}$Zr,
$^{141}$Pr, $^{144}$Sm and $^{208}$Pb are shown in Fig.~3b. For
$^{208}$Pb the parametrization given in Eq.~(3) was
used to calculate the observed energy dependence of $J_I$ (solid
line in Fig.~3b). As expected a strong mass dependence of the volume
integrals is observed, since the strength of the imaginary potential
is quite different for a doubly magic and a strongly deformed
nucleus. Therefore for an energy of $E_{CM} = 10$ MeV, $J_I$ values
can range between 10 and 60 MeV fm$^3$.
\medskip
\noindent For unstable nuclei the mass densities necessary for the
calculation
of the folding potential cannot be obtained from electron scattering
data. In these cases the densities can be calculated in the
$(\sigma\omega\rho)$ model [15,16] used in the relativistic mean-field
theory. We used the parameter set NLSH which is suited for neutron
and proton rich nuclei [17,18,19]. We found that for stable tin and
samarium isotopes the densities calculated in this model compare well
with the experimental data [12]. In order to calculate the
alpha-nucleus potentials for unstable nuclei, the strengths of the
potentials were adjusted to reproduce the parametrized volume
integrals given above.
\bigskip
\noindent {\bf 4. Application to ``p-process'' isotopes}
\medskip
\noindent The alpha-nucleus potentials determined with the folding
procedure are necessary for the calculation of $(\gamma,\alpha)$
photo disintegration cross sections in the p-process. As an example
we consider the inverse reaction $^{144}$Sm$(\alpha,\gamma)^{148}$Gd.
This reaction determines the ratio $^{142}$Nd/$^{144}$Nd in some
meteorites [20]. In previous work the astrophysical S-factors and
photonuclear reaction rates have been generated using the statistical
theory of nuclear reactions as employed by Michaud and Fowler [21].
They used an equivalent square well [ESW] deduced from a Woods-Saxon
potential. However, the effective radius parameter for this ESW is
quite uncertain [22]. For the two different ESW-radii for the
$\alpha$-$^{144}$Sm potential $(R_\alpha = 8.75$ fm and 8.01 fm), the
calculated cross sections differ by a factor of ten.
\midinsert
\centerline{Table 1: S-factors and reaction rates for
$^{144}$Sm$(\alpha,\gamma)^{148}$Gd}
$$
 \offinterlineskip  \tabskip=0pt
 \vbox{
\halign {\strut
\tabskip=0pt
# \hfill
&\vrule# \enskip
&\hfill # \hfill
&\vrule# \enskip
&\hfill # \hfill
&\vrule# \enskip
&\hfill # \hfill
&\vrule# \enskip
&\hfill # \hfill
\tabskip=0pt  \cr
& &ESW & &ESW & &Woods-Saxon & &Folding \cr
& &R = 8.75 fm & &R = 8.01 fm & &potential [24] & &potential$^a$) \cr
\noalign{\hrule\hrule}
S-factor & &$2.3 \cdot 10^{28}$ $^b$) & &$2.3 \cdot
10^{27}$ & & $1.2 \cdot 10^{28}$ & & $7.8 \cdot 10^{27}$ \cr
$(E_\alpha = 9.5$ MeV)~~~[MeV $\cdot$ b] & & & & & & & & \cr
\noalign{\hrule}
reaction rate & &$3.75 \cdot 10^{-15}$ & &$3.72 \cdot 10^{-16}$ &
&$1.95 \cdot 10^{-15}$ & &$1.27 \cdot 10^{-15}$ \cr
$T_9 = 2.5$~~~~~~~~[cm$^3$ mol$^{-1}$ s$^{-1}$] & & & & & & & &  \cr
\noalign{\hrule}
reaction rate & &$2.35 \cdot 10^{-12}$ & &$2.58 \cdot 10^{-13}$ &
&$1.22 \cdot 10^{-12}$ & &$7.56 \cdot 10^{-13}$ \cr
$T_9 = 3.0$~~~~~~~~[cm$^3$ mol$^{-1}$ s$^{-1}$] & & & & & & & &  \cr
}}$$
\noindent $\overline{^a)~\hbox{this work}}$

\noindent $^b)~\hbox{There is a misprint in [22] giving this
value as $2.3 \cdot 10^{29}$.}$
\endinsert
\medskip
\noindent In table 1 we list the astrophysical S-factor  at the Gamow
energy $E_\alpha = 9.5$ MeV and the reaction rates $T_9 = 2.5$ and
$T_9 = 3.0$. In the first two columns the results of calculations
using ESW [22] are given. In the third and fourth column the results
of Hauser-Feshbach calculations using the code SMOKER [23] are shown.
In the first case (column 3) an energy-independent Woods-Saxon
potential with $V = 185$ MeV, $W = 25$ MeV, $R = R_W = 1.4 \cdot
A^{1/3}$ fm and $a = a_W = 0.52$ fm [24], in the second case (column
4) for the real part a folding potential  $(\lambda
= 1.1573)$ and for the imaginary part a Woods-Saxon potential $(W = 10$
MeV, $R_W = 1.4 \cdot A^{1/3}$ fm, $a_W = 0.52$ fm) was used.
\medskip
\noindent With our improved folding potential the reaction rates
shown in the last column of table 1 are about 1/3 of the value for
an ESW with a radius of 8.75 fm. This corresponds about to
1/3 of the ``recommended'' value for the reaction rate in [22] giving
a similar reaction rate as case C in table 1 of [22]. Therefore,
our $^{146}$Sm/$^{144}$Sm ratio is about 0.22 which is consistent
with the cosmochemical data of 0.1 -- 0.7 [20].

\medskip
\noindent The astrophysical S-factor at $E_\alpha = 9.5$ MeV
calculated in the direct-capture model gives an upper limit of about
$10^{22}$ MeV $\cdot$ b, which is more than five orders of magnitude
smaller than the Hauser-Feshbach result.
\bigskip
\noindent {\bf Acknowledgements}
\medskip
\noindent We would like to thank the Deutsche Forschungsgemeinschaft
(DFG), the Fonds zur F\"orderung der wissenschaftlichen Forschung
in \"Osterreich (project P8806--PHY)
and the Austrian-Hun\-ga\-rian Exchange program (project A1).
One of us (T.R.) wants to
thank the Alexander-von-Humboldt Stiftung.
\bigskip
\noindent {\bf References}
\item{1.)} A.M. Kobos, B.A. Brown, R. Linsay, and R. Satchler, Nucl.
Phys. {\bf A425}, 205 (1984)
\item{2.)} H. Oberhummer and G. Staudt, in Nuclei in the Cosmos, ed.
by H. Oberhummer (Springer, Heidelberg, 1991), p. 29
\item{3.)} H. Abele and G. Staudt, Phys. Rev. {\bf C47}, 742 (1993)
\item{4.)} H. Abele, Ph.D. thesis, Univ. of T\"ubingen, 1992
\item{5.)} P. Mohr, H. Abele, V. K\"olle, G. Staudt, H. Oberhummer,
and H. Krauss, Z. Phys. {\bf D}, in press
\item{6.)} G. Raimann, B. Bach, K. Gr\"un, H. Herndl, H. Oberhummer,
S. Engstler, C. Rolfs, H. Abele, R. Neu, and G. Staudt, Phys. Lett.
{\bf B249}, 191 (1990)
\item{7.)} H. Herndl, H. Abele, G. Staudt, B. Bach, K. Gr\"un, H.
Scsribany, H. Oberhummer, and G. Raimann, Phys. Rev. {\bf C44}, R952
(1991)
\item{8.)} T. Rauscher, K. Gr\"un, H. Krauss, H. Oberhummer, E. Kwasniewicz,
Phys. Rev. {\bf C45}, 1996 (1992)
\item{9.)} P. Mohr, H. Abele, R. Zwiebel, G. Staudt, H. Krauss, H.
Oberhummer, A. Denker, J.W. Hammer, and G. Wolf, Phys. Rev. {\bf
C48}, 1420 (1993)
\item{10.)} H. Oberhummer, H. Krauss, K. Gr\"un, T. Rauscher, H.
Abele, P. Mohr, and G. Staudt, Z. Phys. {\bf D}, in press
\item{11.)} S.E. Woosley and W.M. Howard, Ap. J. Suppl. {\bf 36}, 285
(1978)
\item{12.)} H. de Vries, C.W. Jager, and C. de Vries, At. Data and
Nucl. Data Tables {\bf 36}, 495 (1987)
\item{13.)} C. Mahaux, H. Ngo, and G.R. Satchler, Nucl. Phys. {\bf
A449}, 354 (1986); {\bf A456}, 134 (1986)
\item{14.)} C. Mahaux, P.F. Bartignon, R.A. Broglia, and C.H. Dasso,
Phys. Rep. {\bf 120}, 1 (1985)
\item{15.)} P.G. Reinhard, Rep. Prog. Phys. {\bf 52}, 439 (1989)
\item{16.)} Y.K. Gombhis, P. Ring, A. Thienst, Ann. Phys. (N.Y.) {\bf
511}, 129 (1990)
\item{17.)} M.M. Sharma, G.A. Lalazissis, P. Ring, Phys. Lett. {\bf
B317}, 9 (1993)
\item{18.)} M.M. Sharma, P. Ring, Phys. Rev. {\bf C46}, 1715 (1992)
\item{19.)} M.M. Sharma, M.A. Nogarjau, P. Ring, Phys. Lett. {\bf
B312}, 377 (1993)
\item{20.)} A. Prinzhofer, D.A. Papanastassiou, G.A. Wasserburg, Ap.
J. (Letters) {\bf 344}, L81 (1989)
\item{21.)} G. Michaud, W.A. Fowler, Phys. Rev. {\bf C2}, 2041
(1970); Ap. J. {\bf 173}, 157 (1972)
\item{22.)} S.E. Woosley, W.M. Howard, Ap. J. {\bf 354}, L21 (1990)
\item{23.)} F. Thielemann, code SMOKER, unpublished
\item{24.)} F.M. Mann, HEDL-TME 78-83 (1978)
% Literatur zu Zr90
\item{25.)} B.D. Watson, D. Robson, D.D. Talbert, R.H. Davis,
Phys. Rev. {\bf C4}, 2240 (1971)
\item{26.)} L.W. Put, A.M.J. Paans,
Nucl. Phys. {\bf A291}, 93 (1977)
\item{27.)} D.A. Goldberg, S.M. Smith, G.F. Burdzik,
Phys. Rev. {\bf C10}, 1362 (1974)
% Literatur zu Pb208
\item{28.)} J.S. Lilley, M.A. Franey, Da Hsuan Feng,
Nucl. Phys. {\bf A342}, 165 (1980)
\item{29.)} L.L. Rutledge, J.C. Hiebert,
Phys. Rev. {\bf C13}, 1072 (1976)
\item{30.)} V. Corcalciuc, H. Rebel, R. Pesl, H.J. Gils,
J. Phys. {\bf G9}, 177 (1983)
\item{31.)} D.A. Goldberg, S.M. Smith, H.G. Pugh, P.G. Roos, N.S. Wall,
Phys. Rev. {\bf C}, 1938 (1973)
% Literatur zu Ge70
% steht schon bei Zr90
%\item{99.)} B.D. Watson, D. Robson...
\item{32.)} J.B.A. England, S. Baird, D.H. Newton, T. Picazo, E.C.Pollacco,
G.J.Pyle, P.M. Rolph, J. Alabau, E. Casal, A. Garcia,
Nucl. Phys. {\bf A388}, 573 (1982)
\item{33.)} U. Fister, R. Jahn, P. von Neumann-Cosel, P. Schenk, T. K. Trelle,
D. Wenzel, U. Wienands,
Phys. Rev. {\bf C42}, 2375 (1990)
% Literatur zu Sm144/Pr141
\item{34.)} E. Gadioli, E. Gadioli-Erba, P. Guazzoni, L. Zetta,
Phys. Rev. {\bf C37}, 79 (1988)
\item{35.)} U. Baer, H.C. Griffin, W.S. Gray,
Phys. Rev. {\bf C3}, 1398 (1973)
\item{36.)} T. Ichihara, H. Sakaguchi, M. Nakamura, T. Noro, H. Sakamoto,
H. Ogawa, M. Yosoi, M. Ieiri, N. Isshiki, Y. Takeuchi, S. Kobayashi,
Phys.Rev. {\bf C35}, 931 (1987)
\item{37.)} E. M\"uller-Zanotti, to be published

\end